\documentclass[epsf,oneside,12pt]{article}
\usepackage[T1]{fontenc}
\input psfig.sty
\usepackage{amsmath,eufrak}
\include{amsfonts}

\newcommand{\beq}[2]{\begin{equation}#1\label{#2}\end{equation}}
\newcommand{\ceq}[1]{(\ref{#1})}
\newcommand{\mbd}[1]{\mbox{\bf #1}}

\newcommand{\bx}{\mbd{x}}
\newcommand{\by}{\mbd{y}}

\newcommand{\bj}{\mbd{j}}
\newcommand{\bJ}{\mbd{J}}
\newcommand{\bA}{\mbd{A}}
\newcommand{\bB}{\mbd{B}}
\newcommand{\bC}{\mbd{C}}

\newcommand{\bxi}{\BF{\xi}}

\newfont{\mbld}{cmbx10 scaled 800}
\newfont{\cab}{cmsy10 scaled 1200}
\newfont{\scab}{cmsy10 scaled 1000}
\newfont{\bcall}{cmbsy10 scaled 1200}

\newcommand{\BF}[1]{\mbox{\boldmath $#1$}}          
\newcommand{\nablab}{\BF{\nabla}}
\begin{document}
\title{On Abelian Multi-Chern-Simons Field Theories}
\author{Franco Ferrari\\
{\it Institute of Physics, University of Szczecin, ul. Wielkopolska 15,}\\
{\it 70-451 Szczecin, Poland}\thanks{e-mail:
ferrari@univ.szczecin.pl}.}


\maketitle
\renewcommand{\baselinestretch}{1.5}

\abstract{
In this paper a
class of multi-Chern-Simons field theories which
is relevant  to the statistical mechanics of polymer systems is investigated.
Motivated by the problems which one encounters in the treatment of these
theories, a general procedure is presented to eliminate the Chern-Simons
fields from their
action. In this way it has been possible to derive an expression
of the partition function of topologically linked polymers which depends
explicitly on the topological numbers and does not have intractable
nonlocal terms as it happened in previous approaches.
The new formulation of multi-Chern-Simons field theories
is then used to remove and clarify
some inconsistencies and ambiguities which  apparently affect
field theoretical models of
topologically linked polymers.
Finally, the limit of disentangled polymers is discussed.
}
\vfill\eject
\pagestyle{plain}
\section{Foreword}
In this paper we study abelian
multi-Chern-Simons field theories
\cite{Wen:uk,Frohlich:wb,Lee:fw,Birmingham:1991ty,
Wesolowski:1993yk}
coupled to
charged scalar fields.
Models of this kind find a natural application in the statistical
mechanics of  closed polymer rings
 subjected to topological constraints
\cite{Ferrari:ts}
and are
relevant in the phenomenology of the fractional quantum
Hall effect \cite{Cornalba:1997gh}.
An important feature of  these theories is that they exhibit
the phenomenon of charge confinement, which
occurs via a topological mechanism explained in
\cite{Cornalba:1997gh,deWildPropitius:1997wu}.
The aim of this work is to solve
some problems, discussed below,
which arise in the field theoretical formulation
 of polymer systems.

Let us suppose that there are $N$ polymers with trajectories
$P_1,\ldots,P_N$ in a dilute solution. The topological state of the system
is specified in such a way that each trajectory $P_i$ winds up
around trajectory $P_j$ a number $m_{ij}$ of times, $i>j=1,\ldots,N$.
In principle,
it would be desirable to write down an expression of the polymer partition
function which depends on the Gauss linking
 numbers $m_{ij}$'s. In practice, however, it is only
possible to compute the partition function in the space of the Fourier
conjugated
variables $\lambda_{ij}$'s \cite{FKLnpol}.
All attempts to go back to the space of the real topological numbers
$m_{ij}$ 
with
an inverse Fourier transformation led so far to  field theories with
intractable non-local terms \cite{FKLnpol, tanaka}.
Even in the investigation of
the simplest case in which all the polymers
are disentangled, i.e. $m_{ij}=0$ for $i,j=1,\ldots,N$, one encounters
tremendous complications, which can be overcome only resorting to mean-field
like approximations \cite{brerevilgis}.
Moreover, it is also
hard to give a physical meaning to the parameters
 $\lambda_{ij}$. 
At a first sight, in multi-Chern-Simons polymer models
they play the role of
coupling constants and determine the ``strength'' of the 
topological interactions which are necessary to keep the polymer trajectories
in the given topological state.
However, 
there is some freedom in the choice of the
domain in which these parameters are defined, so that
this interpretation cannot be correct.
For example, each $\lambda_{ij}$ can be defined in the interval
$[-\pi,\pi]$, but any other interval of the kind $[l\pi,(l+2)\pi]$,
where $l$ is an arbitrary integer, is also allowed due to the properties
of Fourier transformations. Clearly, from the point of
view of field theories, it makes a great difference if $l=1$ or if
$l>>1$. Furthermore, also the range $(-\infty,+\infty)$
is justified if one starts from a model of open polymers and then
recovers the limit of closed polymers requiring that the ends of the
polymers coincide.

Unfortunately, it is not an easy task
to obtain a reasonably simple expression of the polymer partition
function in terms of the physical parameters $m_{ij}$ and to remove
the above ambiguities.
One problem is that topological interactions in polymer systems
are governed by
 Chern-Simons fields and the contribution of these fields to the partition
function is hard to evaluate. First of all, it is not possible to proceed
perturbatively, since the values of the ``coupling constants'' $\lambda_{ij}$
are not small. On the other side, the action of Chern-Simons is intrinsecally
defined in three dimensions, so that it is difficult
to exploit techniques like the $\epsilon-$expansion which require
its
extension to arbitrary dimensions \cite{Chaichian:1998tf}.
Also from the point of view of numerical
 simulations the situation is not better, since
the lattice formulation of Chern-Simons field theories
encountered so far many obstacles \cite{Bietenholz:2002mt,Berruto:2000dp}.

In order to solve these problems, 
we rewrite the action of
multi-Chern-Simons field theories
with the help of
a Hubbard-Stratonovich
transformation and successively eliminate the Chern-Simons fields.
The resulting field theories
contain only scalar fields and
a set of auxiliary
fields which have no dynamics. 
The procedure used is quite general, although it has been developed
for the special multi-Chern-Simons field theories which are relevant to
polymers.
Exploiting the new formulation of the polymer partition function,
we are able to prove that there are not
ambiguities in the field theoretical description of polymer systems,
because all the possible domains in which the coupling constants
$\lambda_1,\ldots,\lambda_\Gamma$
may be
defined lead to equivalent models. Finally,
we clarify the physical meaning of these coupling constants: They play the
role of Lagrange multipliers. The constraints which they impose are explicitly
computed.

The material presented in this paper is divided as follows.
In Section 2 we consider an action with two Chern-Simons
fields coupled with multiplets of charged scalar fields which are 
invariant
under a $U(n_1)\times U(n_2)$ group of global symmetry.
The relations of this model with the statistical mechanics of two
topologically linked polymers have been
discussed in details in \cite{Ferrari:ts,Ferrari:yz} and
are here only briefly summarized.
Starting from this simple example of
multi-Chern-Simons field theories,
we illustrate the problems which arise in polymer models.
In Section 3 it is shown how it is possible to eliminate
the Chern-Simons fields using a suitable Hubbard-Stratonovich transformation.
In this way we obtain an expression of the partition
function of polymers in terms of the topological numbers $m_{ij}$
in which the action is polynomial in the fields. In previous approaches,
instead, the action contained non-polynomial and non-local terms.
In Section 4
we check that the
non-uniqueness  of the domain of integration of the coupling constants
disappears
due to a symmetry which was hidden in the original formulation in terms
of Chern-Simons fields.
Finally, in Sections 5 and 6
these results are generalized to the case of multi-Chern-Simons
field theories describing the statistical mechanics of an arbitrary
number of polymers.

\section{The Two-Polymers Model}
Let us consider the action:
\begin{eqnarray}
S(\lambda)&=&
\int d^3x\left[ 
\imath\kappa\bB\cdot(\nablab\times \bA)+
|(\nablab-\imath\kappa\bB)\Psi_1|^2+m_1^2|\Psi_1|^2\right]  
\nonumber\\
&+&\int d^3x\left[|(\nablab-\imath\lambda\bA)\Psi_2|^2+m_2^2|\Psi_2|^2\right]
\label{acttwo}
\end{eqnarray}
In Eq.~\ceq{acttwo} $\imath=\sqrt{-1}$ and
the symbols $\Psi_i,\Psi_i^*$, $i=1,2$, denote
multiplets of charged fields:
\beq{\Psi_i=(\psi_i^1,\ldots,\psi_i^{n_i})\qquad\qquad\qquad
\Psi_i^*=(\psi_i^{*1},\ldots,\psi_i^{*n_i})}{repfie}
In our notations:
\begin{eqnarray}
|(\nablab-\imath\kappa\bB)\Psi_1|^2&=&\sum_{a=1}^{n_1}(\nablab+\imath\kappa\bB)
\psi^{*a}_1 \cdot(\nablab-\imath\kappa\bB)\psi_1^a\label{notone}\\
|(\nablab-\imath\kappa\bA)\Psi_2|^2&=&\sum_{a=1}^{n_2}(\nablab+\imath\kappa\bA)
\psi^{*a}_2 \cdot(\nablab-\imath\kappa\bA)\psi_2^a\label{nottwo}\\
|\Psi_i|^2&=&\sum_{a=1}^{n_i}\psi_i^{*a}\psi_i\qquad\qquad i=1,2\label{notthr}
\end{eqnarray}
It is easy to check by rescaling the field $\bB$
that the Chern-Simons coupling constant $\kappa$ is irrelevant and that only
the coupling constant $\lambda$ appears in the physical amplitudes of the
theory. 

Field theories such as those of Eq.~\ceq{acttwo} enter in various
physical problems. Here we consider the case of two closed polymers
with trajectories $P_1$ and $P_2$ and lengths $L_1$ and $L_2$
respectively.  The
trajectories are
constrained to satisfy the following topological constraint:
\beq{\chi(P_1,P_2)=m\qquad\qquad\qquad m=0,\pm 1,\pm 2\ldots}{topcond}
where $\chi(P_1,P_2)$ is the Gauss linking number given by:
\beq{\chi(P_1,P_2)=\frac
1{4\pi}\int_0^{L_1}ds_1\int_0^{L_2}ds_2\dot{\bx}_1(s_1)\cdot\left[\dot{\bx}_2
(s_2)
\times \frac{(\bx_1(s_1)-\bx_2(s_2))}{|\bx_1(s_1)-\bx_2(s_2)|^3}
\right]}{glndef}
In the above equation $P_1$ and $P_2$ are represented 
by two closed curves $\bx_1(s_2),\bx_2(s_2)$, where
$s_1$ and $s_2$ are the arc lengths of the trajectories.
In terms of the bond currents
\beq{\bj_i(\bx)=\int_0^{L_i}ds_i\bx_i(s_i)\delta^{(3)}(\bx-\bx_i(s_i)),}
{curdef}
Eq.~\ceq{glndef} may be rewritten as follows:
\beq{m=\frac 1{4\pi}\int d^3x d^3y\bj_1(\bx)\cdot\left[\bj_2(\by)
\times\frac{(\bx-\by)}{|\bx-\by|^3}\right]}{currtopcond}
It is possible to show that, in the Lorentz gauge, in which the fields $\bA$
and $\bB$ are completely transverse,
 the partition function of this two-polymer
system coincides with the following amplitude \cite{Ferrari:yz,Ferrari:eq}:
\beq{Z=
\int_0^{2\pi}\frac{d\lambda}{2\pi} e^{-\imath m\lambda}Z(\lambda)}{parfunone}
where
\beq{Z(\lambda)=
\int{\cal
D}(fields)|\psi_1^1(\bx)|^2|\psi_2^1(\by)|^2e^{-S_2(\lambda)} 
}{parfuntwo}
and
\beq{{\cal D}(fields)=\int{\cal D}\bA{\cal
D}\bB\prod_{i=1}^2\prod_{a_i=1}^{n_i} \left[{\cal D}\psi_i^a{\cal
D}\psi_i^{*a} \right]}{meas}
To make contact with polymer physics, we should keep in mind that it is
still necessary to continue analytically the partition function
$Z$ to the limit of zero replica numbers $n_1$ and $n_2$.
Moreover, one should also add to the action \ceq{acttwo} the so-called
excluded volume interactions, which
take into account the steric repulsions of the monomers.
However, both
 zero replica limit and excluded volume interactions are
 irrelevant in the present context and
will be ignored.

We see from Eq.~\ceq{parfunone} that one has
to consider the sum over the partition functions
$Z(\lambda)$ for all values of the coupling constant $\lambda$ in the
interval $[0,2\pi]$.
This is a consequence of the fact that
the topological condition \ceq{topcond} has been imposed by inserting in the
partition function the $\delta$ of Kronecker $\delta_{m,\chi(P_1,P_2)}$,  which
in the Fourier representation is given by:
\beq{
\delta_{m,\chi(P_1,P_2)}=
\int_0^{2\pi}\frac{d\lambda}{2\pi}
e^{-\imath\lambda(m-\chi(P_1,P_2))}}{fourep}
This integration over $\lambda$
is a further complication with respect to standard
field theories, which makes it difficult to study the physical
properties
 of the two-polymer model given above.
For example, let us note that the right
hand side of Eq.~\ceq{fourep} is invariant under
the shifts:
\beq{\lambda\longrightarrow\lambda+\pi k\qquad\qquad\qquad k=0,
\pm1,\pm2\ldots}{lamshi}
Therefore, for consistency,
also the
partition function $Z$
should be invariant as a function of $\lambda$ under these shifts.
However, such invariance is not evident from
Eqs.~(\ref{parfunone}--\ref{parfuntwo})
and from the action \ceq{acttwo}.
Other difficulties arise
if we wish to describe the behavior of two disentangled
polymer rings starting from the partition function of
Eqs.~(\ref{parfunone}--\ref{parfuntwo}). As a matter of fact,
even in the limit of zero topological number $m$,
the integration over $\lambda$ remains complicated.
In general, the investigation of the $m=0$ limit
is  problematic in models of topologically linked polymers 
based on the Edwards approach. The only concrete results
have been achieved up to now
in the case of dense solutions, where mean field-like
approximations are possible \cite{brerevilgis}.

In principle, one can easily eliminate the coupling constant
$\lambda$ in \ceq{parfunone} by performing a simple Gaussian
integral, but the new partition function contains non-local
operators whose treatment by analytical methods is difficult
\cite{FKLnpol}.
On the other side, it is possible to study
the partition function $Z(\lambda)$
by means of  field theoretical techniques.
However, the knowledge of the properties of $Z(\lambda)$ does not
provide a very deep insight into
the properties
of the final partition function $Z$.
\section{Elimination of the Chern-Simons Fields}
A more transparent formulation of the two-polymer problem, in which
the role of the couplingg constant $\lambda$ is explicit, can be
provided by means of two Hubbard-Stratonovich transformations.
As a first step, let us rewrite the partition function of
Eqs.~(\ref{parfunone}--\ref{parfuntwo}) in the following way:
\begin{eqnarray}
Z&=&\int_0^{2\pi}\frac{d\lambda}{2\pi}
e^{-\imath m\lambda}\int{\cal D}\bA
{\cal D} \bB\prod_{i=1}^2\left[
\prod_{a_i=1}^{n_i}{\cal
D}\psi_i^{a_i}{\cal D}\psi_i^{*a_i}\prod_{b_i=1}^{n_i}
{\cal D}\BF{\xi}_i^{b_i}{\cal D}\BF{\xi}_i^{*b_i} \right]\nonumber\\
&&|\psi_1^1(\bx)|^2|\psi_2^1(\by)|^2
e^{-S_0-S_1(\lambda)}
\label{newparfun}
\end{eqnarray}
where $\BF{\xi}^{*a}_2, \BF{\xi}^a_2$  represent
two sets of auxiliary complex vector fields and $S_0, S_1(\lambda)$
are given by:
\begin{eqnarray}
S_0&=&
\sum_{i=1}^2
\sum_{a_i=1}^{n_i}
\int d^3x\left\{
-\imath
\left[
\nablab\psi_i^{*a_i}
\cdot
\BF{\xi}_i^{a_i}+
\nablab\psi_i^{a_i}\cdot\BF{\xi}_i^{*a_i}\right]\right.\nonumber\\
&+&\left.m_i^2|\Psi_i^{a_i}|^2+\BF{\xi}_i^{a_i}
\cdot \BF{\xi}_i^{*a_i}\right\}\label{szero}
\end{eqnarray}
\beq{
S_1(\lambda)=
\imath\kappa\int d^3x 
\bB\cdot(\nablab\times \bA)+\imath\kappa\int d^3x\bB\cdot\bJ_1+
\imath\lambda\int d^3x\bA\cdot\bJ_2}{suno}
Here we have introduced the vector fields
\beq{\bJ_i=\frac1\imath\sum_{a_i=1}^{n_i}\left[\psi_i^{*a_i}\BF{\xi}_i^{a_i}
-\psi_i^{a_i}\BF{\xi}_i^{*a_i}\right]}{curralg}
which are related to the total matter currents of the replica fields
$\psi_i^{a_i},\psi_i^{*a_i}$.
This connection with matter currents becomes more explicit if we
consider the classical equations of motion of the fields
$\bxi_i^{a_i},\bxi_i^{*a_i}$:
\begin{eqnarray}
\bxi_1^{a_1}=(\nablab-\imath\kappa\bB)\psi_1^{a_1}&\qquad\qquad&
\bxi_1^{*a_1}=(\nablab+\imath\kappa\bB)\psi_1^{*a_1}\label{equmoto}\\
\bxi_2^{a_2}=(\nablab-\imath\kappa\bA)\psi_2^{a_2}&\qquad\qquad&
\bxi_2^{*a_2}=(\nablab+\imath\kappa\bA)\psi_2^{*a_2}\label{equmott}
\end{eqnarray}
Substituting Eqs. \ceq{equmoto} and \ceq{equmott} in \ceq{curralg} one
obtains:
\begin{eqnarray}
\bJ_1&=&\frac
1\imath\sum_{a_1=1}^{n_1}\left[\psi_1^{*a_1}(\nablab-\imath\kappa\bB)
\psi_1^{a_1} -\psi_1^{a_1}(\nablab+\imath\kappa\bB)
\psi_1^{*a_1}\right]\label{ocuraftequmot}\\
\bJ_2&=&\frac
1\imath\sum_{a_2=1}^{n_2}\left[\psi_2^{*a_2}(\nablab-\imath\lambda\bA)
\psi_2^{a_2} -\psi_2^{a_2}(\nablab+\imath\lambda\bA)
\psi_2^{*a_2}\right]\label{tcuraftequmot}
\end{eqnarray}
which are exactly the total abelian matter currents of the replica
fields.

At this point we are ready to show that the partition function
\ceq{parfunone} and \ceq{newparfun} are equivalent.
To prove that, it is sufficient to perform in Eq.~\ceq{newparfun} the
change of variables:
\beq{
\bxi_1^{a_1}
=\bxi_1^{\prime
a_1}+\imath(\nablab-\imath\kappa\bB)\psi_1^{a_1}\qquad\qquad
\bxi_1^{*a_1}
=\bxi_1^{\prime *
a_1}+\imath(\nablab+\imath\kappa\bB)\psi_1^{* 1}}{trasimo}
\beq{\bxi_2^{a_2}
=\bxi_2^{\prime
a_2}+\imath(\nablab-\imath\lambda\bA)\psi_2^{a_2}
\qquad \qquad 
\bxi_2^{*a_2}
=\bxi_2^{\prime *
a_2}+\imath(\nablab+\imath\lambda\bA)\psi_2^{* 2}}{trasimt}
After this substitution, the result is exactly the partition function
of Eqs.~(\ref{parfunone}--\ref{parfuntwo}), apart from an irrelevant
constant coming from the Gaussian
integration over the decoupled primed fields $\bxi_i^{\prime
a_i},\bxi_i^{*\prime a_i}$, $i=1,2$, $a_i=1,\ldots,n_i$.
Hubbard--Stratonovich transformations of this kind are common in
polymer physics \cite{vilgisrep,kle}.
They can be rewritten in a more familiar form in
terms of real vector fields, coinciding with the real and imaginary
parts of the fields $\bxi_i^{a_i}$ and $\bxi_i^{*a_i}$. This point
will be discussed in more details in the Appendix.

It is now easy to integrate out the Chern-Simons field from the
partition function \ceq{newparfun}. To this purpose, we have to
consider the path integral:
\beq{Z_{\psi\mbld{\xi}}=\int{\cal D}\bA{\cal D}\bB e^{-S_1(\lambda)}
}{auxtwofie}
A first integration over the $\bB$ fields gives:
\beq{Z_{\psi\mbld{\xi}}=\int{\cal D}\bA e^{-\lambda\int
d^3x\bA\cdot\bJ_2}
\delta(\nablab\times\bA+\bJ_1)
}
{abparfun}
The $\delta-$function in Eq.~\ceq{abparfun} enforces the constraint:
\beq{\nablab\times\bA+\bJ_1=0
}{constone}
which implies that the current $\bJ_1$ is conserved as expected.
As a matter of fact, taking the divergence of both members of
Eq.~\ceq{constone}, one obtains: $\nablab\cdot\bJ_1=0$.

Solving the constraint \ceq{constone} with respect to $\bA$,
Eq.~\ceq{abparfun} becomes:
\beq{Z_{\psi\mbld{\xi}}
=\exp\left[
-\imath\frac\lambda{4\pi}\int
d^3xd^3y\left(\nablab\frac1{|\bx-\by|}\times\bJ_2(\by) \right)\cdot
\bJ_1(\bx)\right]}{abpfsolved}
The substitution of \ceq{abpfsolved} in the partition function
\ceq{newparfun} gives the following result:
\begin{eqnarray}
Z&=&\int_0^{2\pi}\frac {d\lambda}{2\pi}e^{-\imath m\lambda}
\int{\cal D}(fields)'
|\psi_1^1(\bx)|^2|\psi_2^1(\by)|^2\nonumber\\
&\times&\exp
\left[-S_0
-\imath\frac\lambda{4\pi}\int
d^3xd^3y\left(\nablab\frac1{|\bx-\by|}\times\bJ_2(\by) \right)\cdot
\bJ_1(\bx)\right]
\label{phyforparfun}\end{eqnarray}
The field integration measure is now:
\beq{{\cal D}(fields)'=
\prod_{i=1}^2\prod_{a_i=1}^{n_i}
{\cal D}\psi_i^{a_i}
{\cal D}\psi_i^{*a_i}{\cal D}\bxi_i^{a_i}
{\cal D}\bxi_i^{*a_i}
}{phyformea}
\section{The Limit of Disentangled Polymers}
With the new formulation of the partition function given by
Eq.~\ceq{phyforparfun}, the role of the parameter $\lambda$ in the
two-polymer model and its invariance under
the shifts $\lambda\rightarrow\lambda+\pi$
have become transparent.
As a matter of fact, performing the simple integration over $\lambda$
one obtains:
\begin{eqnarray}
Z&=&
\int{\cal D}(fields)'
 |\psi_1^1(\bx)|^2
|\psi_2^1(\by)|^2
e^{-S_0}\nonumber\\
&\times&
\delta\left(\textstyle m-\frac\lambda{4\pi}\int
d^3xd^3y\bJ_1(\bx)\cdot\left(
\bJ_2(\by)\times\nablab\frac1{|\bx-\by|}\right)
\right)\label{confinpar}
\end{eqnarray}
where the measure ${\cal D}(fields)'$ and the action $S_0$ have been
given in Eqs.~\ceq{phyformea} and \ceq{szero} respectively.
Clearly, the constraint
\beq{
\textstyle \frac\lambda{4\pi}\int
d^3xd^3y\bJ_1(\bx)\cdot\left(
\bJ_2(\by)\times\nablab\frac1{|\bx-\by|}\right)
=m}{connt}
is the analogous of the present field theoretical formalism of the
topological constraint \ceq{currtopcond}. 

In the limit $m=0$, the
partition function \ceq{phyforparfun} becomes:
\begin{eqnarray}
Z_{m=0}&=&
\int{\cal D}(fields)'
 |\psi_1^1(\bx)|^2
|\psi_2^1(\by)|^2
e^{-S_0}\nonumber\\
&\times&
\delta\left(\textstyle \frac\lambda{4\pi}\int
d^3xd^3y\bJ_1(\bx)\cdot\left(
\bJ_2(\by)\times\nablab\frac1{|\bx-\by|}\right)
\right)\label{zmzero}
\end{eqnarray}
From Eq.~\ceq{zmzero} it turns our that topological interactions do
not vanish if $m=0$ as one could naively expect from the fact that the
polymers are disentangled in this case.
The reason is that the topological interactions are still necessary
when the polymers get too near at some point in order to prevent the
crossing of the trajectories, which would modify the value of $m$.

Eq.~\ceq{phyforparfun} shows also that the form of the
partition function $Z$ does
not change under a shift of the coupling constant $\lambda$ of the
kind given in Eq.~\ceq{lamshi}.
As a matter of fact, let us perform the shift $\lambda\longrightarrow\lambda+
\pi k=\lambda'$, $k=\pm1,\pm2,\ldots$
in Eq.~\ceq{parfunone}, so that we get the new partition function:
\beq{
Z_{shifted}=\int_{\pi k}^{\pi(k+1)}
\frac{d\lambda'}{2\pi}e^{-im\lambda'}
Z(\lambda')=(-1)^{km}\int_0^{2\pi}\frac{d\lambda}{2\pi}e^{-\imath m\lambda}
Z(\lambda+\pi k)
}{zshift}
Repeating the same steps which led from Eq.~\ceq{parfunone} to
Eq.~\ceq{phyforparfun}, we obtain:
\begin{eqnarray}
Z_{shifted}&=&(-1)^{km}\int_0^{2\pi}\frac{d\lambda}{2\pi}e^{-\imath m\lambda}
\int{\cal D}(fields)'
 |\psi_1^1(\bx)|^2
|\psi_2^1(\by)|^2
e^{-S_0}\nonumber\\
&\times&
\exp
\left[
-\imath\frac{\lambda+\pi k}{4\pi}\int
d^3xd^3y\left(\nablab\frac1{|\bx-\by|}\times\bJ_2(\by) \right)\cdot
\bJ_1(\bx)\right]\label{zshifdwa}
\end{eqnarray}
The integration over $\lambda$ imposes once again the constraint
\ceq{connt} in the partition function $Z_{shifted}$. As a consequence,
it is easy to show that
\beq{Z_{shifted}=Z}{zshiide}
for $k=0,\pm1,\pm2\ldots$, because of the relation
\beq{
e^{-\imath\frac{\pi k}{4\pi}\int
d^3xd^3y\left(\nablab\frac1{|\bx-\by|}\times\bJ_2(\by) \right)\cdot
\bJ_1(\bx)}=e^{-\imath\pi k m}=(-1)^{km}}{exide}
which is enforced by this constraint.

Thanks to the invariance under the shift \ceq{lamshi} and
Eq.~\ceq{zshiide},
we are also able to prove the following identity:
\beq{\tilde Z\equiv\int_{-\infty}^{+\infty}\frac{d\lambda}{2\pi} e^{-\imath
m\lambda}
Z(\lambda)=Z}{ideimp}
where 
$Z$ is  the partition function given above in (\ref{parfunone}) and
$Z(\lambda)$ is defined in 
Eq.~(\ref{parfuntwo}).
The partition function $\tilde Z$ differs from $Z$ only
by the domain of integration of the parameter $\lambda$,
which in this case is the real line $(-\infty,+\infty)$.
This is what one obtains
if one derives the model of two closed polymers starting from two open
polymers and then requiring that their ends coincide
\cite{Ferrari:ts,FKLnpol}.

To verify Eq.~\ceq{ideimp}, we divide the domain of integration over
$\lambda$ in the infinite number of intervals $[2\pi l,2\pi (l+1)]$,
where $-\infty\le l\le +\infty$ is an integer.
In this way, $\tilde Z$ becomes of the form:
\beq{
\tilde Z=
\sum_{l=-\infty}^{+\infty}\int_{2\pi
l}^{2\pi(l+1)}\frac{d\lambda_l}{2\pi}
 e^{-\imath
m\lambda_l}Z(\lambda_l)}{ggg}
Since $m$ is an integer, it is also possible to write:
\beq{
\tilde Z=
\sum_{l=-\infty}^{+\infty}\int_0
^{2\pi}\frac{d\lambda}{2\pi}
 e^{-\imath
m\lambda}Z(\lambda+2\pi l)}{hhh}
Using Eq.~\ceq{zshifdwa} and the
constraint \ceq{connt} it is now easy to show 
that the terms depending on $l$ factorize as follows: 
\begin{eqnarray}
\bar Z&=&\int_0^{2\pi}\frac{d\lambda}{2\pi} e^{-\imath
m\lambda}\int{\cal D}(fields)'
 |\psi_1^1(\bx)|^2
|\psi_2^1(\by)|^2
e^{-S_0}\nonumber\\
&\times&
\delta\left(\textstyle \frac\lambda{4\pi}\int
d^3xd^3y\bJ_1(\bx)\cdot\left(
\bJ_2(\by)\times\nablab\frac1{|\bx-\by|}\right)
\right)
\sum_{l=-\infty}^{+\infty}e^{2\pi \imath ml}
\label{gggfin}
\end{eqnarray}
Comparing the above expression with the expression of the partition
function $Z$ of Eq.~\ceq{confinpar} it is possible to conclude that
$\tilde Z = Z$ apart from the infinite constant factor
$\sum_{l=-\infty}^{+\infty}e^{2\pi \imath ml} =
\sum_{l=-\infty}^{+\infty} 1$, which does not change the physics
of the problem.
\section{The $N$-Polymer Model}
In this Section we consider the extension of the two-polymer model
discussed above
to the more realistic case of the fluctuations of $N$ polymers.
Let us denote with $P_1,\ldots,P_N$ the trajectories of the $N$
polymers, which  are constrained to satisfy the following relations:
\beq{\chi(P_i,P_j)=m_{ij}\qquad\qquad 
\begin{array}{rcl}
m_{ij}&=&0,\pm 1,\pm 2,\ldots\\
i,j&=&1,\ldots,N
\end{array}}{multiconst}
These conditions can be enforced in the partition function which
describes
the statistical mechanics of the $N$ polymers by inserting the
following product of Kronecker delta's:
\beq{
\prod_{i=2}^N\prod_{j=1\atop j<1}^{N-1}\delta(\chi(P_i,P_j)-m_{ij})=
\prod_{i=2}^N\prod_{j=1\atop j<1}^{N-1}\int_0^{2\pi}\frac{d
\lambda_{ij}}{2\pi} e^{-\imath\lambda_{ij}(m_{ij}-\chi(P_i,P_j))}}
{fourexp}
where the $\lambda_{ij}$'s are elements of a $N\times N$ matrix of
Fourier parameters such that $\lambda_{ij}=0$ whenever $j\ge i$ for
$i,j=1,\ldots N$.
The field theoretical version of the $N-$polymer model has been
derived in \cite{FKLnpol}. Its partition function is given by:
\beq{{\cal Z}_N=
\int\prod_{i=2}^n\prod_{j=1\atop j<i}^{N-1}\frac{d\lambda_{ij}}{2\pi}
e^{-\imath m_{ij}\lambda_{ij}} {\cal Z}_N(\lambda_{ij})}
{partnpol}
where
\beq{{\cal Z}_N(\lambda_{ij})=\int{\cal D}(\bA\bB\Psi_i\Psi_i^*)\prod_{i=1}^N
\psi_i^1(\bx^i)\psi^{*1}_i(\by^i)e^{-{\cal S}}}{partlamnpol}
and the action $\cal S$ is:
\beq{{\cal S}=\imath\kappa\sum_{i=1}^{N-1}\int
d^3x\bA^i\cdot(\nablab\times\bB^i)+\sum_{i=1}^N\int
d^3x\left[|(\nablab-\imath\bC^i)
\Psi_i|^2+m_i^2|\Psi_i|^2\right]}{actn}
The fields $\Psi_i,\Psi^*_i$, $i=1,\ldots,N$, represent multiplets of
replica fields:
\beq{\Psi_i=(\psi_i^1,\ldots,\psi_i^{n_i})\qquad
\Psi_i=(\psi_i^{*1},\ldots,\psi_i^{*n_i})}{nmulti}
and the vector fields $\bC^i$ are linear combinations of the
Chern-Simons fields $\bA^i,\bB^i$:
\beq{
\bC^i=\sum_{j=1}^{i-1}\lambda_{ij}\bA^j(1-\delta_{i1})+\bB^i\delta_{i1}}
{ciajbj}
Finally,
\beq{{\cal D}(\bA\bB\Psi_i\Psi_i^*)=\prod_{i=1}^{N-1}{\cal D}\bA^i
{\cal D}\bB^i\prod_{j=1}^N\prod_{a_j=1}^{n_j}
{\cal D}\psi_j^{a_j}{\cal D}\psi_j^{*a_j}}{dddfields}
\section{Elimination of the Chern-Simons Fields from the $N-$Polymer
Model}
First of all, let us introduce auxiliary complex fields $\bxi_i^{a_i},
\bxi_i^{*a_i}$, where $i=1,\ldots,N$ and $1\le a_i\le n_i$.
In a similar way as we did in the case $N=2$ in Section 3, it is now
possible to rewrite partition function \ceq{partnpol} in the following
way:
\beq{{\cal Z}_N(\lambda_{ij})=\int{\cal D}(
\bA\bB\Psi_i\Psi_i^*\bxi_i\bxi_i^*)e^{-{\cal S}_0-{\cal
S}_1(\lambda_{ij})}}
{zniwithxi}
where:
\beq{{\cal D}(
\bA\bB\Psi_i\Psi_i^*\bxi_i\bxi_i^*)=
\prod_{i=1}^{N-1}{\cal D}\bA^i
{\cal D}\bB^i\prod_{j=1}^N\prod_{a_j=1}^{n_j}
{\cal D}\psi_j^{a_j}{\cal D}\psi_j^{*a_j}{\cal D}\bxi_j^{a_j}{\cal
D}\bxi_j^{*a_j}}{nppmeas}
\beq{{\cal S}_0=\sum_{i=1}^N\sum_{a_i=1}^{n_i}\int d^3x\left[
-\imath\left(\nablab\psi_i^{*a_i}\cdot\bxi_i^{a_i}+\nablab\psi_i^{a_i}\cdot
\bxi_i^{*a_i}\right)+\bxi_i^{a_i}\cdot\bxi_i^{*a_i}\right]}{nszero}
\beq{
{\cal S}_1(\lambda_{ij})=\imath\kappa\sum_{i=1}^{N-1}\int d^3x
\bA^i\cdot(\nablab\times\bB^i)+\imath\sum_{i=1}^N\bC^i\cdot\bJ_i}{nsone}
and
\beq{\bJ_i=\frac1\imath\sum_{a_i=1}^n\left[\psi_i^{*a_i}\bxi_i^{a_i}+
\psi_i^{a_i}\bxi_i^{*a_i}\right]}{npj}
At this point we are ready to perform the integration over the fields
$\bA^i,\bB^i$ in the partition function ${\cal Z}_N$. To this purpose,
we need to evaluate the path integral:
\beq{
{\cal Z}_{N\psi\bxi}=\int\prod_{i=1}^N{\cal D} \bA^i{\cal D}\bB^i
e^{-{\cal S}_i(\lambda_{ij})}}{npauxparf}
One finds
after a first integration over the fields $\bB^j$'s:
\beq{
{\cal Z}_{N\psi\bxi}=\int\prod_{i=1}^N{\cal D}\bA^i
e^{-\imath\kappa\sum_{i=2}^N\sum_{j=1}^{i-1}\lambda_{ij}\bA^j\cdot\bJ_i}
\prod_{i=1}^{N-1}\delta(\kappa\nablab\times\bA^i-\bJ_i)}{auxaftela}
The product of Dirac $\delta-$functions in \ceq{auxaftela} enforces
the constraints:
\beq{
\kappa\nablab\times\bA^i=\bJ_i}{nncons}
whose solution (in components) is:
\beq{
A_\mu^i=\frac1{4\pi\kappa}\int
d^3y\epsilon_{\mu\nu\rho}\frac{(x-y)^\nu} {|\bx-\by|^3}J_i^\rho(\by)
\qquad\qquad\mu,\nu,\rho=1,2,3}{solnpol}
It is now easy to show that:
\beq{
{\cal Z}_{N\psi\bxi}
=\exp\left\{
-\frac{\imath}{4\pi}\sum_{i=2}^N\sum_{j=1}^{i-1}\int d^3x
d^3y
\lambda_{ij} \epsilon_{\mu\nu\rho}
\frac{(x-y)^\nu}
{|\bx-\by|^3}J_i^\mu(\bx)J_j^\rho(\by)\right\}}{solconn}
Substituting this result in the expression of the partition function
${\cal Z}_N$ if Eq.~\ceq{partnpol} one obtains:
\begin{eqnarray}
{\cal Z}_N&=&\int\prod_{i=2}^N\prod_{j=1\atop
j<i}^{N-1}\frac{d\lambda_{ij}} {2\pi} e^{-\imath
m_{ij}\lambda_{ij}} \prod_{i=1}^N\prod_{a_i=1}^{n_i}
{\cal D}\psi_i^{a_i}{\cal D}\psi_i^{*a_i}{\cal D}\bxi_i^{a_i}{\cal
D}\bxi_i^{*a_i} 
\nonumber\\
&&\!\!\!\!\!\!\!\!\!\!\!\!\!\!\!\!\!\!\!\!\!\!\!\!\!
\exp\left\{
-{\cal S}_0+\frac \imath{4\pi}
\sum_{i=2}^N\sum_{j=1}^{i-1}\int d^3x
d^3y
\lambda_{ij} \bJ_i(\bx)\cdot\left[
\nablab\frac1{|\bx-\by|}\times\bJ_j(\by)\right]\right\}
\label{newnparfun}
\end{eqnarray}
From the formulation of the partition function ${\cal Z}_N$ given by
Eq.~\ceq{newnparfun} it turns out that the Fourier variables
$\lambda_{ij}$ play the role of Lagrange multipliers imposing the
constraints:
\beq{m_{ij}=\frac \imath{4\pi}\int d^3x
d^3y
\lambda_{ij} \bJ_i(\bx)\cdot\left[
\nablab\frac1{|\bx-\by|}\times\bJ_j(\by)\right]}{nnconnn}
which
are the generalization to the $N$ polymer case  of condition
\ceq{connt}.
Once again, in the limit in which all polymers are disentangled, the
effects of the topological interactions do not disappear.
The invariance of the partition function ${\cal Z}_N$ with respect to
the shifts:
\beq{\lambda_{ij}\rightarrow\lambda_{ij}+\pi k_{ij}\qquad\qquad
k_{ij}=0,\pm1,\pm2,\ldots}{nlamshi}
can also be proved using similar methods as those employed in Section 4.
\section{Conclusions}
In this paper we have studied abelian multi-Chern-Simons field theories
coupled with matter fields. The attention has been concentrated to those models
which are relevant to the statistical mechanics of polymers, but some
results are valid for any theory containing Chern-Simons fields.
Motivated by the difficulties
which one encounters when dealing with these theories
due to the presence of
the topological fields and their kinetic terms, 
 we have proposed a procedure to eliminate these fields from the
action.
The advantage is that now
the partition function explicitly
depends on the topological numbers $m_{ij}$ and the polymer action is 
polynomial in the fields, contrarily to the action derived in
 \cite{FKLnpol}.
The price to be paid is that the new action contains auxiliary
vector fields and has a non-local two-body interaction.

With the help of the new formulation it has been possible to show that
the action \ceq{actn} 
is invariant under the shifts of the coupling
constants $\lambda_{ij}$ given in
 Eq.~\ceq{nlamshi}.
This symmetry,
which was not a priori evident in the action
\ceq{actn}, has been used in order to show
the equivalence of all  models of topologically
linked polymers differing by the range of integration
of the Fourier variables $\lambda_{ij}$'s, see
Eqs.~\ceq{zshiide} and \ceq{ideimp}.
The generalization of these results to any $N$ starting from
the partition function of Eq.~\ceq{newnparfun} is straightforward.
Finally, it has been clarified that the parameters $\lambda_{ij}$'s in polymer
models are Lagrange multipliers, which impose the
contraints \ceq{nnconnn}. These conditions represent
 clearly the field theoretical
version of the topological constraints \ceq{topcond}, but are not topological
relations by themselves. This is a natural consequence of the fact that,
in the process of elimination of the Chern-Simons fields,
topological and non-topological terms have been mixed together.

To conclude, we would like to address some problems which are still open.
First of all, experiments suggest that in the presence
of topological constraints there are attractive forces acting on polymers
\cite{levene}.
In particular,  there are evidences that
 the strength of these forces increases
with increasing complexity of the topological configuration of the system.
A perturbative calculation at the one-loop approximation confirms the
presence of such forces in the two-polymer model \cite{Ferrari:yz} ,
but it is difficult to estimate how their strength depend on $m$ starting
from the partition function \ceq{confinpar}.
Another open question is how the phenomenon of confinement
that is active in multi-Chern-Simons field theories may influence the
statistical behavior of the polymers.

%

\section{Appendix A: Hubbard-Stratonovich Transformations}
In deriving Eqs.~(\ref{newparfun}--\ref{suno}),
as well as Eqs.~(\ref{zniwithxi}--\ref{nsone}),
 we have used the
generalization to path integral of the following Gaussian integral
formula:
\begin{eqnarray}
\int\prod_{\alpha=1}^ndz_\alpha\bar
z_\alpha\exp\left[-\sum_{\alpha,\beta=1}^n\bar z_\alpha
A_{\alpha\beta}z_\beta+\imath \sum_{\alpha=1}^n(\bar b_\alpha
z_\alpha+b_\alpha\bar z_\alpha)\right]&=&\nonumber\\
=(2\pi \imath)^n(\det
A)^{-1}\exp\sum_{\alpha,\beta =1}^n\left[-\bar b_\alpha
(A^{-1})_{\alpha \beta}b_\beta\right]&\mbox{}&\label{gauintfor}
\end{eqnarray}
In the above equation $z_\alpha,\bar z_\alpha$ are a set of complex
variables and $b_\alpha,\bar b_\alpha$ are
constants. To prove Eq.~\ceq{gauintfor} it is sufficient to perform
the transformation
\begin{eqnarray}
z_\alpha=z'_\alpha+\imath(A^{-1})_{\alpha\beta}b_\beta\label{trans1}\\
\bar z_\alpha=\bar z'_\alpha+\imath\bar b_\beta(A^{-1})_{\alpha\beta}
\label{trans2}
\end{eqnarray}
which is the analogous of Eqs.~(\ref{trasimo}--\ref{trasimt}).
The Gaussian formula \ceq{gauintfor} can be brought in an
equivalent form after switching to real components $x_1,\ldots,x_{2n}$
and $c_1,\ldots,c_{2n}$ of $z_\alpha$ and $b_\alpha$ respectively:
\begin{eqnarray}
z_\alpha=x_\alpha+\imath x_{n+\alpha}&\qquad\qquad& \bar
z_\alpha=x_\alpha-\imath x_{n+\alpha}\label{prez1}\\
b_\alpha=c_\alpha+\imath c_{n+\alpha}&\qquad\qquad& \bar
b_\alpha=c_\alpha-\imath c_{n+\alpha}\label{prez2}
\end{eqnarray}
Substituting Eqs.~(\ref{prez1}--\ref{prez2}) in \ceq{gauintfor} one obtains
the following Gaussian identity, which is familiar in polymer physics because
it is used to simplify the excluded volume interactions:
\beq{
\int\prod_{j=1}^{2n}[dx_j]\exp\left(-\sum_{j=1}^{2n}
x_j^2+2\imath\sum_{j=1}^{2n}c_jx_j\right)=
(2\pi)^n\exp\left(-2\sum_{j=1}^{2n}c_j^2
\right)}{secgaufor}

\end{document}